\documentclass[twocolumn]{aastex63}
\submitjournal{ApJ (accepted August 12, 2020)}

\shorttitle{IR separation between thin-shelled O/C-AGB stars}
\shortauthors{Lewis et al.}
\graphicspath{{./}{figures/}}

\begin{document}

\title{Infrared color separation between thin-shelled oxygen-rich and carbon-rich AGB stars}

\correspondingauthor{Megan O. Lewis}
\email{melewis@unm.edu}

\author[0000-0002-8069-8060]{Megan O. Lewis}
\affiliation{University of New Mexico, Department of Physics and Astronomy \\
Albuquerque, NM 87131, USA}
\affiliation{National Radio Astronomy Observatory \\
Array Operations Center\\
Socorro, NM 87801, USA}

\author[0000-0003-0615-1785]{Ylva M. Pihlstr\"om}
\altaffiliation{Ylva M.\ Pihlstr\"om is also an Adjunct Astronomer at the\\ National Radio Astronomy Observatory.}
\affiliation{University of New Mexico, Department of Physics and Astronomy \\
Albuquerque, NM 87131, USA}

\author[0000-0003-3096-3062]{Lor\'ant O. Sjouwerman}
\affiliation{National Radio Astronomy Observatory \\
Array Operations Center\\
Socorro, NM 87801, USA}

\author[0000-0002-9390-955X]{Luis Henry Quiroga-Nu\~nez}
\altaffiliation{Luis-Henry Quiroga-Nu\~nez is a Jansky Fellow of the National Radio Astronomy Observatory.}
\affiliation{National Radio Astronomy Observatory \\
Array Operations Center\\
Socorro, NM 87801, USA}
\affiliation{University of New Mexico, Department of Physics and Astronomy \\
Albuquerque, NM 87131, USA}



\begin{abstract}
We present 43 GHz VLA spectra for 51 AGB sources with the goal of verifying an infrared (IR) color cut intended to separate Carbon-rich (C) and Oxygen-rich (O) AGB sources throughout the Galaxy. The color cut is a simple line in the $[K_s]-[A]$ versus $[A]-[E]$ color-color diagram based on 2MASS and MSX photometry, and was originally derived from SiO detection rates in the Bulge Asymmetries and Dynamical Evolution (BAaDE) sample. The division is fully supported by the spectra presented here, which show that SiO maser detections lie on the O-rich side, and SiO nondetections and a single HC$_3$N detection are found on the C-rich side of the division. We further compare the color cut with classifications of the sources based on Low-Resolution Spectra (LRS) from the Infrared Astronomical Satellite (IRAS), and find good agreement, verifying that the division is a reliable and efficient method for differentiating O- and C-rich AGB sources. These observations also demonstrate that single lines detected in the BAaDE survey around 42.9 GHz are almost certainly the $^{29}$SiO v=0 line. SiO maser sources where this rare isotopologue transition is brighter than the dominant $^{28}$SiO lines have not been reported before, and our observations show that these sources can reverse their behavior such that the typical ratios of $^{28}$SiO and $^{29}$SiO are restored within a few years.

\end{abstract}



 \section {Introduction}
 Asymptotic Giant Branch (AGB) stars, evolved stars between 0.8 and 8 $M_\sun$, are characterized by large circumstellar envelopes (CSEs). They can be divided into two main chemical types, Oxygen-rich (O) and Carbon-rich (C), based on the C/O ratio of the envelope. These two categories show different spectral energy distributions (SEDs) and spectral signatures in multiple wavebands (e.g., \citealt{lloydevans}) because excess C or O atoms bond with other metals in the CSE and create an envelope dominated by either C-bearing or O-bearing molecules, respectively. Identifying the chemical types of AGB sources has been, perhaps, most successful at infrared (IR) wavelengths, as shown by multiple surveys involving low-resolution spectra (LRS; \citealt{olnon86}; \citealt{kvb97}). However, some misidentifications and ambiguous cases exist (i.e., incomplete or featureless spectra, C-stars with silicate emission features), and reliable identifications can only be made for AGB stars whose IR emission does not suffer too much from interstellar and circumstellar extinction. 
 \par Efficient identification of C- and O-rich AGB sources is useful for reliably counting C-rich AGB sources in the Galaxy, which would inform about processes like dredge-up in stellar evolution models and the enrichment and return of material to the interstellar medium from low- and intermediate-mass evolved stars. Confirmed C-rich AGB stars in the bulge are particularly rare \citep{matsunaga17}, and identifying candidates in this category is essential to develop a complete picture of the age and metallicity of the bulge as well as the the formation of C-rich AGB sources. 
 \par Radio maser signals from OH, H$_2$O, and especially SiO are common in O-rich AGB stars and can be used as a diagnostic tracer of these stars. These maser signals are often very bright and can be used to identify O-rich AGB stars that are significantly affected by interstellar extinction in the optical and IR, including stars that lie deep in the Galactic bulge \citep{messineo02}. Although no masers of comparable brightness or prevalence are found in C-rich AGB stars, many thermal molecular transitions can be seen at radio wavelengths in C-rich AGB stars given enough sensitivity. 
 There have been many observations that focus on single, nearby, bright C-rich AGB sources. For example, IRC+10216 has been examined in great detail (eg.\,\citealt{dvtlim08}; \citealt{ten10}; \citealt{chau12}; \citealt{he19}). 
 These studies show that C-stars in particular have a rich chemistry which gives rise to multiple spectral signatures at radio wavelengths.

 \par Much larger samples, as well as an efficient method of classifying an AGB star as either C- or O-rich, are required in order to determine distributions of C-/O-rich objects in the Galaxy, classify large numbers of C-rich AGB sources, and identify molecules that are common for a large set of typical C-AGB sources. Investigations that consider samples of AGB stars often collect their source-lists from preexisting catalogues from the literature, and are thereby forced to gather their C and O stars from separate sources (\citealt{ortiz05}; \citealt{lumsden02}; \citealt{suh11}). 
 This precludes comparing numbers of C and O AGB stars systematically and homogeneously.
 The most uniform analyses of AGB chemistry are based on IR LRS. \cite{kvb97} classified $\sim$5000 Infrared Astronomical Satellite (IRAS) sources using this method. IR LRS however are limited to sources that are about 10 Jy or brighter at 12 and 25 $\mu$m \citep{1988iras....1.....B} and therefore this method cannot be easily extended to populations in the Galactic bulge.

 \par Photometric methods of dividing C- and O-rich objects have the capability of classifying thousands of sources more efficiently than spectroscopic methods. IR-color cuts based on IRAS \citep{vdvH88}, Midcourse Space eXperiment (MSX; \citealt{ortiz05}), Akari \citep{ishihara11}, and combined 2 Micron All Sky Survey (2MASS) and MSX (\citealt{lumsden02}; \citealt{lewis20}) photometry have been proposed to separate large numbers of C- and O-rich AGB stars. A quick and powerful method of spectroscopically confirming these cuts is necessary before large scale studies of these objects are possible. In particular, spectral radio signals from C-rich AGB stars near the transitions of SiO masers (43 and 86 GHz) would be a useful diagnostic to categorize AGB sources. In depth spectral studies of AGB stars are sparse at 43 GHz as most molecular surveys are conducted at mm wavelengths (e.g.\,\citealt{bujarrabal94}; \citealt{decin18}; \citealt{he19}). 
 
 \par The results presented in this paper are from follow-up observations of C-rich candidate sources covered by the one-epoch Bulge Asymmetries and Dynamical Evolution (BAaDE) survey (\citealt{michael19}; \citealt{trapp18}; Sjouwerman et al.\,2020, in preparation). The BAaDE project is a targeted SiO maser survey of $\sim$ 28,000 IR-selected sources, primarily evolved stars likely to harbor SiO masers \citep{lewis20}. The survey covers the full Galactic plane, utilizing both the NSF's Karl G. Jansky Very Large Array (VLA) at 43 GHz and the Atacama Large Millimeter/submillimeter Array (ALMA) at 86 GHz.  The spectral setups allow not only for multiple SiO transitions to be observed, but also for the detection of a set of C-bearing molecular transitions. This is of interest in terms of separating, and characterizing, the C-rich versus O-rich AGB populations, as the BAaDE IR selection was based on MSX colors and could not separate between the two \citep{scc09}. The BAaDE targets were selected to be thin-shelled AGB sources from an MSX equivalent \citep{scc09} of IRAS color-color region IIIa as defined by \cite{vdvH88}; this means that the BAaDE sample is quite different from samples which rely on optical and OH/IR sources (\citealt{lumsden02}; \citealt{ortiz05}; \citealt{suh11}). \cite{lewis20} found that the following zero-magnitude corrected IR color-color division (not corrected for extinction) was a promising method for separating these thin-shelled O-rich and C-rich AGB stars:
 \begin{equation}
     [K_s]-[A]=4.5([A]-[E])-1.2, 
 \end{equation} 
 where  $[K_s]-[A]$ values greater (redder) than this boundary indicate a likely C-rich AGB source, and smaller (bluer) values are likely O-rich. The 2MASS $[K_s]$, MSX $[A]$, and MSX $[E]$ bands are centered at wavelengths of 2.19, 8.28, and 21.34 $\mu$m and zero-magnitude corrections of 55.810, 8.666, and 665 Jy are applied respectively so that comparisons and colors can be constructed across instruments.
 \par Establishing this division was partly based on a small set of ALMA detections of C-bearing molecular lines confirming the C-rich chemistry and the position of these sources in the color-color diagram. The VLA observations did not reveal similar direct observational proof of a C-rich CSE.  However, the BAaDE observations were designed to detect bright maser lines, and did not provide sufficient integration time for regular detections of weaker, thermal carbon-bearing lines. Here we present a deeper integration of a subset of the BAaDE sources, with the goal of verifying the C- and O-rich IR separation method.
 
\par In Sect.\,\ref{obs} we describe the selection and observation of our sample of candidate C-rich AGB sources. Section \ref{results} covers the spectral line detections made in the sample, and the clear distinction between the spectra of O-rich AGB sources and candidate C-rich AGB sources. In Sect.\,\ref{discussion} we discuss the evidence in support of the IR-C/O division and possible limitations that arise. Also in Sect.\,\ref{discussion}, we draw attention to a population of sources with anomalously bright SiO isotopologue lines. Our conclusions are summarized in Sect.\,\ref{conclusion}, and the spectra from the follow-up observations appear in the Appendix. 
 
 \section{Observations}\label{obs}  
 We observed 51 sources, all of which were previously observed as part of the BAaDE project, with the aim of positively identifying C-stars and firmly establishing the IR color division between C- and O-rich AGB sources. 
 \subsection{Source selection} 
\begin{figure}[htb]
     \centering
     \includegraphics[width=.45\textwidth]{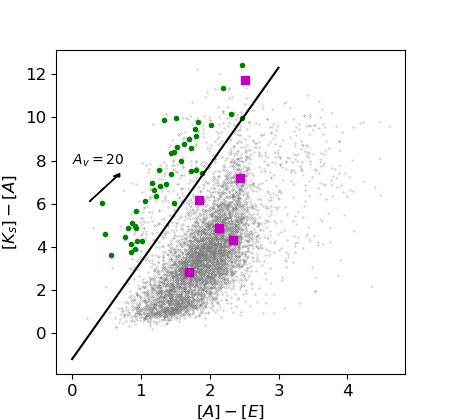}
     \caption{Color-color diagram of BAaDE sources published in \cite{lewis20} (grey points) including sources selected for follow-up based on their color (green dots) or based on their BAaDE spectra (magenta squares). One source (\textit{ad3a-07012}) selected by its spectra does not have a reliable [E]-magnitude and so does not appear on this plot. The diagonal line represents the division given by Equation 1 \citep{lewis20} where sources to the upper left of the line have colors indicative of C-rich AGB stars and sources to the lower right are likely O-rich; hence, all of the color-selected sources are above this line. The black arrow shows the reddening vector, corresponding to 20 magnitudes of visual extinction, from \cite{lumsden02} which assumes silicate features typical of O-rich dust and that extinction goes as $\lambda^{-1.75}$.}
     \label{fig:selection}
 \end{figure}

In \cite{lewis20} C- and O-rich stars are separated empirically in a color-color diagram. The O-rich region was directly confirmed by the high rate of SiO maser detection, while the VLA portion of the BAaDE survey was not designed to detect weaker lines of C-bearing molecules which may be expected in C-rich CSEs (\citealt{dvtlim08}; \citealt{chau12}). Thus, to verify this division, we performed follow-up observations of two types of sources: one group falling into the anticipated C-rich region of the color-color diagram which had no line detections, and one group falling mainly into the anticipated O-rich color-color region with a single line detection that could not easily be ascribed to a specific CSE chemistry. The selected sources are highlighted in the [K$_s$]$-$[A] versus [A]$-$[E] diagram in Fig.\,\ref{fig:selection}.
 
\subsubsection{Sources with a single detected line near 42.9 GHz}\label{sourcesline}
Seven of our targets were selected because they showed an ambiguous single-line detection near 42.9 GHz during the BAaDE survey. The BAaDE spectra for these sources can be found on the BAaDE survey website\footnote{\label{website}\url{http://www.phys.unm.edu/~baade/specs/table.shtml}} and their names and coordinates are listed in Table \ref{tab:singlines}. The nearest SiO line to 42.9 GHz is the v=0 $J=1-0$ transition of $^{29}$SiO. As $^{28}$SiO is the more common isotopologue, one may not expect to see any $^{29}$SiO lines without the $^{28}$SiO v=1 or $^{28}$SiO v=2 line. However, upon inspection of $\sim$14,000 BAaDE spectra from the survey website the $^{29}$SiO v=0 line can be brighter than the $^{28}$SiO lines within a single source in a few rare instances (where confusion between isotopologue transitions is eliminated for sources with multiple lines detected). Therefore, it is possible for $^{29}$SiO v=0 to be the only detection in a source. HC$_7$N has a transition near 42.9 GHz as well (within 17 MHz or $\sim$110 km\,s$^{-1}$ of the $^{29}$SiO line). This line is also not expected to be the brightest at these frequencies as the HC$_5$N J=16$-$15 transition at 42.602 GHz is dominant in the C-rich AGB stars CIT 6 and IRC+10216 \citep{chau12}. In short, a single line at this frequency is not expected in either O-rich or C-rich AGB sources as other transitions are more likely to be dominant in this frequency range in either case. In the new, higher sensitivity observations we searched for multiple lines with the goal of identifying the originally detected line. The IR colors of these sources suggest five are O-rich, one C-rich, and one cannot be classified in the scheme of \cite{lewis20} due to the lack of a reliable [E]-magnitude; all of these BAaDE spectra are ambiguous. 

\newpage
\subsubsection{Sources chosen by color}\label{sourcescolor}
Forty-four potential candidates for C-AGB stars were chosen from the BAaDE survey based on their IR colors and magnitudes. These were selected from the $\sim$400 sources whose colors suggest they may be C-stars that were covered by the VLA portion of the BAaDE survey \citep{lewis20}. Selected sources are brighter than magnitude 3 in MSX [A]-band, and fall on the C-rich side of the color cut in Equation 1. Sources selected under these criteria are most likely to be bright candidate C-rich AGB stars. Thermal C-bearing molecular lines are expected to be much weaker than SiO masers which is why the selection includes the magnitude cut ($[A]<3$), intended to limit the sample to nearby sources in the absence of better distance estimates. A cross-matched BAaDE-\textit{Gaia} sample was established after these observations were complete; however, all but two sources in our sample still lack reliable distance estimates because of the absence of \textit{Gaia} counter-parts or large uncertainties in parallaxes (see Sect.\,\ref{results:color}). Circumstellar and interstellar extinction are likely responsible for the limited \textit{Gaia} information available on these sources (Quiroga-Nu\~nez et al.\,2020, submitted). 
A discussion of local BAaDE sources with known distances is presented in Quiroga-Nu\~nez et al. (2020, submitted), which shows that only 6\% of the entire BAaDE catalogue can be attributed to sources within 2 kpc of the Sun with good parallaxes, and that the BAaDE-\textit{Gaia} sources are mostly moderately luminous ($\sim$1500 L$_\sun$), very thin-shelled AGB stars that are atypical by the criteria in this paper.
\par All of these sources were nondetections in the BAaDE survey. A nondetection of SiO in BAaDE does not necessarily guarantee that the source is not O-rich, as the SiO maser intensity varies greatly as a function of the stellar phase. Even among samples of known SiO maser sources redetection rates are often less than 80\% because of the effects of variability (\citealt{ylva}; \citealt{michael18}). Incidentally, this study also supports an $\sim$80\% redetection rate as 5 out of 6 (83\%) sources that showed reliable lines during the BAaDE survey had one or more lines detected in this follow-up (see Sect.\,\ref{results}). Apart from variability, an SiO nondetection would also be expected if C is more abundant than O, and in the new, deeper observations we searched the spectra for molecular detections that would confirm this. 

\begin{deluxetable*}{ccccccccc}
\tablecaption{Observation dates and results for sources chosen for their single-line detections in the BAaDE survey. All observations were made in January and February 2020. IRAS LRS classifications are taken from \cite{kvb97} and descriptions are therein. E classifications are likely O-rich sources and other classifications do not aid in differentiating O- and C-rich AGB sources. All line-of-sight velocities derived from the follow-up observations have uncertainties of $\pm$7 km\,s$^{-1}$ because of the spectral resolution of the observations.\label{tab:singlines}}
\tablehead{
\colhead{BAaDE} & \colhead{RA} & \colhead{Dec} & \colhead{IRAS} & \colhead{BAaDE} & \colhead{Follow-up} & \colhead{IR} & \colhead{LRS} & \colhead{V$_{\mathrm{LOS}}$}
\\ 
\colhead{name} & \colhead{(J2000)} & \colhead{(J2000)} & \colhead{name} & \colhead{date} & \colhead{result} & \colhead{region} & \colhead{class} & \colhead{km\,s$^{-1}$}
}

\startdata
ad3a-01808 & 17:43:05.400 & $-$30:18:02.88 & IRAS 17398$-$3016 & Feb 2016 & $^{28}$SiO  v=0, 1, 2, 3 & O & I & $-$17\\
 &  & & & & $^{29}$SiO v=0 & & & \\
ad3a-07012 & 18:07:12.696 & $-$22:40:47.28 & no match & March 2013 & $^{28}$SiO v=1, 2 & - & - & 92 \\
 & & & & & & & & \\
ad3a-10509 & 18:30:40.728 & $-$10:44:26.52 & no match & March 2017 & $^{28}$SiO  v=0 & O & - & 140 \\
 & & & & & & & & \\
ad3a-11939 & 18:43:07.248 & $-$06:42:39.60 & IRAS 18404$-$0645 & March 2017 & $^{28}$SiO  v=0, 1, 2 & O & E & 93\\
 &  & & & & $^{29}$SiO v=0 & & & \\
ad3a-13236 & 18:42:24.888 & $-$02:51:57.24 & IRAS 18397$-$0254 & March 2017 & no detection & O & E & - \\
 & & & & & & & &\\
ce3a-00109 & 18:02:48.288 & $-$22:19:51.96 & IRAS 17597$-$2219 & March 2013 & $^{29}$SiO  v=0 \tablenotemark{a} & O & I & $-$43\\
 & & & & & & & &\\
ce3a-00142 & 18:34:02.208 & $-$12:06:51.84 & IRAS 18312$-$1209 & March 2017 & unidentified & C & P & 93 or $-$21\tablenotemark{b}\\
\enddata
\tablenotetext{a}{Single line identified using typical velocities associated with Galactic longitude of the source}
\tablenotetext{b}{Velocities for line identifications of $^{29}$SiO and HC$_7$N respectively}
\end{deluxetable*}

\subsection{Data characteristics} 
For the purpose of this work, each of the selected BAaDE sources has been reobserved in follow-up observations using the VLA Q-band receivers. The new observations cover a wider range of frequencies in order to include more potential transitions of carbon-bearing molecules in addition to multiple SiO transitions. 
Data from the BAaDE survey were taken between roughly 42 and 43 GHz for less than one minute per source, reaching a root-mean-square (RMS) noise level of about 20 mJy\,beam$^{-1}$\,channel$^{-1}$.
The new observations reported on here were taken between 41.5$-$43.5 GHz (covering SiO maser and C-bearing molecular transitions) and 45$-$47 GHz (covering only C-bearing transitions) for about 12 minutes per source, reaching an RMS noise level of about 3 mJy\,beam$^{-1}$\,channel$^{-1}$. 
This is the noise level required to detect a line equivalent to the 45.5 GHz HC$_3$N line in IRC+10216 at 3-sigma in one channel out to about $\sim$1.1$-$1.5 kpc. \citep{dvtlim08}. 
The spectral resolution was 1.7 km\,s$^{-1}$ in the BAaDE data and 7 km\,s$^{-1}$ in the follow-up data. 
Data reduction techniques used in the BAaDE survey are described in Sjouwerman et al. (2020, in preparation). 
The follow-up data were reduced using standard phase and amplitude calibration in AIPS and spectra were produced at the phase centers. 
Source information and observing dates are listed in Table \ref{tab:singlines} and Table \ref{tab:ccolor}.

\section{Results}\label{results}
All spectra from the new observations are shown in the Appendix. Spectra from the BAaDE survey were obtained via the survey website\textsuperscript{\ref{website}}. 
The characteristics of the follow-up spectra depend heavily on the criteria under which they were selected. In particular, the spectra of the sources selected for their single-line detections (Sect.\,\ref{sourcesline}) indicate that these sources are mostly O-rich AGB stars, while sources chosen for their colors (Sect.\,\ref{sourcescolor}) are less definitively identified but likely mostly C-rich. The two sets are therefore discussed separately. 

\subsection{Single-line sources} 
Of the seven single-line selected sources described in Sect.\,\ref{sourcesline}, one was undetected, one shows an ambiguous detection, one shows a single thermal SiO line, and four show SiO maser lines in our follow-up observations. The velocities of four of the detected sources (three maser and one thermal) are consistent with the single-line detections from the BAaDE survey being the $^{29}$SiO v=0 transition, keeping in mind that the spectral resolution of the follow-up data is 7 km\,s$^{-1}$. The velocity of \textit{ad3a-07012} may indicate that the BAaDE detection was a rare false positive caused by a noise spike. 
In general we conclude that in most cases the brightest line in the spectra of these sources at the time of the BAaDE observations was, unexpectedly, the $^{29}$SiO v=0 maser line. 
Each source is discussed in more detail below.
\begin{itemize}
    \item \textit{ad3a-01808} displays several SiO transitions: $^{28}$SiO v=0, $^{28}$SiO v=1, $^{28}$SiO v=2, $^{28}$SiO v=3, and $^{29}$SiO v=0 yielding an average line-of-sight velocity of $-$17 km\,s$^{-1}$. This compares well if we assign the $^{29}$SiO v=0 transition to the BAaDE detection, finding $-$16.7 km\,s$^{-1}$ for those observations. In the new observations of this source, the $^{29}$SiO v=0 line is fainter than the $^{28}$SiO v=1, $^{28}$SiO v=2, and $^{28}$SiO v=3 lines, showing that reversed line ratios between $^{29}$SiO and $^{28}$SiO are not a set characteristic of a given AGB star. The strength of the lines from these isotopologues can switch back and forth within an individual source (see also Sect.\,\ref{isovar}). 
    \item \textit{ad3a-07012} shows $^{28}$SiO v=1 and 2 lines during this observation. The velocities associated with these lines (94 and 91 km\,s$^{-1}$ respectively) do not match the velocity of the line in the original 2013 BAaDE data if it is identified as $^{29}$SiO v=0 ($-$247 km\,s$^{-1}$) or HC$_7$N ($-$362 km\,s$^{-1}$). Upon reinvestigating the preliminary BAaDE spectra, the detection is 9-sigma in a single 250 kHz (1.7 km\,s$^{-1}$) channel and may be a noise peak. This is an O-rich source whose maser emission was either off or below the BAaDE detection threshold during the BAaDE observations. The 84 and 73 mJy single channel (1024 kHz or 7 km\,s$^{-1}$) peaks were only significant with the increased sensitivity of the follow-up observation.
    \item \textit{ad3a-10509} shows a tentative detection of the $^{28}$SiO v=0 line. The detection is 6-sigma when three-channel (21 km\,s$^{-1}$) averaging is applied. Being a likely SiO emitter, it makes sense to assign the $^{29}$SiO v=0 transition to the single-line BAaDE detection. The validity of this is strongly supported by the derived line-of-sight velocities ($\sim$140 km\,s$^{-1}$) in both surveys, differing by less than 2 km\,s$^{-1}$. We presume that since 2016 the $^{29}$SiO v=0, and possibly other maser lines, have turned off due to stellar variability.
    \item \textit{ad3a-11939} hosts $^{28}$SiO v=0, $^{28}$SiO v=1, $^{28}$SiO v=2, and $^{29}$SiO v=0 emission with an average line-of-sight velocity of 93 km\,s$^{-1}$. The $^{29}$SiO v=0 line is within 300 kHz ($\sim$2 km\,s$^{-1}$) of its velocity in the 2017 BAaDE observations, confirming the identification of the original transition. The $^{29}$SiO v=0 line is the brightest in the source during both observations. This source is IRAS 18404$-$0645 and is identified as an O-rich AGB star based on the 9.7 $\mu$m emission feature in its LRS.
    \item \textit{ad3a-13236} hosts no detectable lines in the follow-up observations. This source has an IRAS counterpart (IRAS 18397-0254) and is identified as an O-rich AGB source based the  LRS 9.7 $\mu$m emission feature. As stated previously in Sect.\,\ref{sourcescolor}, nondetections can occur in SiO maser sources due to variability. Based on the IRAS LRS, and the findings above for ad3a-01808 and ad3a-10509, we extrapolate that the detection in the 2017 BAaDE data was the $^{29}$SiO v=0 line and that maser emission turned off before our follow-up observations.
    \item \textit{ce3a-00109} shows a single line in the follow-up observations around 600 kHz (4 km\,s$^{-1}$) away from the line detected in the 2013 BAaDE observations (i.e., in complete agreement given the spectral resolution). This is likely the $^{29}$SiO v=0 line making the line-of-sight velocity of the source $-$42 km\,s$^{-1}$. Alternatively, if this is the HC$_7$N line the line-of-sight velocity is $-$160 km\,s$^{-1}$. As this source lies at a Galactic longitude of 7.6$^{\circ}$, where the velocities of BAaDE sources range between about $-$150 to 250 km\,s$^{-1}$, and recalling the previous three cases, we consider the $^{29}$SiO v=0 identification more likely. The IRAS LRS for this source is incomplete.
    \item \textit{ce3a-00142} is the only source chosen for its single-line detection that has IR colors indicative of a C-rich AGB source---although it should be noted that its [K$_s$]$-$[A] value is very high at 11.7 (see Fig.\,\ref{fig:selection}) and it is unclear whether the division extends this far due to the low number of BAaDE sources with similar colors. It shows a single line again in the follow-up observations within 200 kHz (2 km\,s$^{-1}$) of the line detected in the 2017 BAaDE observations. If this line is the $^{29}$SiO v=0 the derived line-of-sight velocity is 93 km\,s$^{-1}$; if it is the HC$_7$N line then the velocity is $-$21 km\,s$^{-1}$. In this case the ambiguity of the single line is not resolved as the velocity range of BAaDE sources at this longitude is about $-$60 to 170 km\,s$^{-1}$. If the line is the $^{29}$SiO v=0 line it does not appreciably undermine the significance of the IR-color division because of the high [K$_s$]$-$[A] value of the source and because of the inherent uncertainties in color values (see Sect.\,\ref{con:redvar}). 
\end{itemize}

Several general statements can be made about the sources chosen under the single-line criteria. The ambiguity of the line near 42.9 GHz in the BAaDE data is resolved. This line is the $^{29}$SiO v=0 line, likely in all reliable cases (i.e., \textit{ad3a-07012} and \textit{ce3a-00142} possibly excluded), and therefore the $^{29}$SiO v=0 line can be the brightest SiO transition in a source. Given the SiO detections in most of the O-rich region sources, we find that they are O-rich AGB sources, which largely validates the color cuts laid out in \cite{lewis20}. 
Two sources also have IRAS LRS classifications that help differentiate O- and C-rich AGB sources, both of which support O-rich identifications \citep{kvb97}. No unambiguous detections of C-bearing molecules were made in these sources. 
\par In addition, these data have demonstrated the variable nature of the relative brightnesses of $^{29}$SiO and $^{28}$SiO lines in the source \textit{ad3a-01808}. In this source, $^{29}$SiO v=0 was the brightest line in the BAaDE data and the $^{28}$SiO v=1,2, and 3 lines are all brighter than the $^{29}$SiO v=0 line in the follow-up data. This behavior has not been reported before and this will be explored in a forth-coming paper. 
\subsection{Color-selected sources} \label{results:color}
Of the 44 sources described in Sect.\,\ref{sourcescolor}, 43 sources were undetected in the follow-up observations and one showed molecular emission. 
\par We detect the thermal $^{28}$SiO v=0 and HC$_3$N v=0 ($J=5-4$) transitions in \textit{ad3a-11235}. The detected HC$_3$N line is outside of the frequency range of the BAaDE data at 45.490 GHz, and both the HC$_3$N and SiO lines are well below the sensitivity of the BAaDE data with single-channel (1024 kHz) peaks at 11 and 8 mJy respectively. The HC$_3$N v=0 ($J=5-4$) line is the brightest line between 35$-$50 GHz in both CIT 6 and IRC+10216 \citep{chau12}. Detecting this line in a C-rich source is therefore more compatible with previous observations than the possibility of detecting the HC$_7$N v=0 ($J=38-37$) line discussed in Sect.\,\ref{sourcesline}. 
\par
\begin{figure*}[htb]
    \centering
    \includegraphics[width=.9\textwidth]{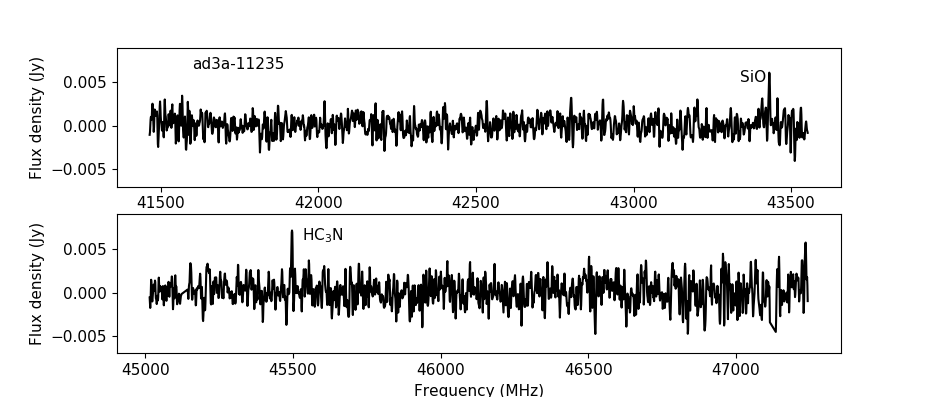}
    \caption{Spectra of source \textit{ad3a-11235} which displays the $^{28}$SiO v=0 line and the HC$_3$N v=0 ($J=5-4$) line. Hanning smoothing is applied across seven channels. The top and bottom panels correspond to different frequency ranges}
    \label{fig:ad3a-11235}
\end{figure*}

 Both detected lines are wider than the few km\,s$^{-1}$ maser detections discussed in the previous section, with the SiO line spanning 21 km\,s$^{-1}$ and the HC$_3$N line spanning 35 km\,s$^{-1}$, where the velocity widths are calculated from the number of channels between the first and last 3-sigma channels inclusively. The spectra of this source is therefore reproduced with seven-channel ($\sim$ 49 km\,s$^{-1}$) Hanning smoothing in Fig.\,\ref{fig:ad3a-11235} to more clearly show these lines. With the peak channel of the SiO line at $-$32 km\,s$^{-1}$ and the  HC$_3$N line at $-$26 km\,s$^{-1}$ the velocities agree to within the errors. This source was also observed by the IRAS and \textit{Gaia} surveys; it is classified as a C-star by its IRAS LRS and several other means (\citealt{MSIII}, \citealt{kvb97}, \citealt{alksnis}, \citealt{chen}), and is 789$\pm$207 pc from the Sun as determined by directly inverting its \textit{Gaia} parallax \citep{gaia}. Of the sources in this sample with \textit{Gaia} parallaxes available, \textit{ad3a-11235} is likely the nearest. \textit{ad3a-09804} is the only other source with a \textit{Gaia} counterpart and a parallax error small enough to obtain a distance with no priors \citep{bailerjones15}, and its inverted \textit{Gaia} parallax yields a distance of 814$\pm$150 pc.

\newpage
\startlongtable
\begin{deluxetable*}{cccccccc}
\tablecaption{Observation dates and results for sources chosen by IR color. All observations were made in January and February 2020. IRAS LRS classifications are taken from \cite{kvb97} and descriptions are therein. C and F classifications are likely C-rich sources, A and E classifications are likely O-rich sources, and other classifications do not aid in differentiating O- and C-rich AGB sources. All line-of-sight velocities derived from the follow-up observations have uncertainties of $\pm$7 km\,s$^{-1}$ because of the spectral resolution of the observations.\label{tab:ccolor}}
\tablehead{
\colhead{BAaDE} & \colhead{RA} & \colhead{Dec} &\colhead{IRAS name} & \colhead{BAaDE} & \colhead{Follow-up} & \colhead{LRS} & \colhead{V$_{\mathrm{LOS}}$}
\\ \colhead{name} & \colhead{(J2000)} &\colhead{(J2000)} &\colhead{} &\colhead{date} & \colhead{result} & \colhead{class} &\colhead{km\,s$^{-1}$}
}

\startdata
ad3a-00345 & 17:48:52.008 & $-$33:57:20.52 & IRAS 17455$-$3356 & Feb 2016 & no detection  & - & - \\
ad3a-03860 & 17:43:04.080 & $-$28:36:58.32 & IRAS 17399$-$2835 & Feb 2016 & no detection  & - & - \\
ad3a-04092 & 17:47:44.736 & $-$28:26:38.40 & no match        & Feb 2016 & no detection  & - & - \\
ad3a-06452 & 18:02:04.248 & $-$23:37:42.96 & IRAS 17590$-$2337 & March 2013 & no detection & P & - \\
ad3a-06564 & 18:00:21.144 & $-$23:22:01.92 & IRAS 17573$-$2322 & March 2013 & no detection & - & - \\
ad3a-06816 & 18:03:39.000 & $-$22:59:40.92 & no match        & March 2013 & no detection & - & - \\
ad3a-07039 & 17:58:09.408 & $-$22:37:39.36 & no match        & March 2013 & no detection & - & - \\
ad3a-09614 & 18:21:39.072 & $-$13:40:54.84 & IRAS 18188$-$1342 & March 2017 & no detection & U & - \\
ad3a-09688 & 18:26:43.728 & $-$13:23:18.96 & IRAS 18238$-$1325 & April 2017 & no detection & F & - \\
ad3a-09711 & 18:20:51.432 & $-$13:17:11.76 & IRAS 18180$-$1318 & March 2017 & no detection & A\tablenotemark{c} & - \\
ad3a-09804 & 18:19:50.592 & $-$12:58:05.88 & IRAS 18170$-$1259 & March 2017 & no detection & F & - \\
ad3a-09912 & 18:21:14.040 & $-$12:32:40.20 & IRAS 18184$-$1234 & March 2017 & no detection & F & - \\
ad3a-11232 & 18:27:34.248 & $-$08:37:21.72 & IRAS 18248$-$0839 & March 2017 & no detection & C & - \\
ad3a-11235 & 18:31:36.360 & $-$08:35:27.24 & IRAS 18288$-$0837 & April 2017 & $^{28}$SiO v=0, HC$_3$N & C & $-$29 \\
ad3a-11311 & 18:29:01.344 & $-$08:17:19.68 & no match        & April 2017 & no detection & - & - \\
ad3a-11329 & 18:27:07.512 & $-$08:13:09.12 & IRAS 18244$-$0815 & March 2017 & no detection & C & - \\
ad3a-11767 & 18:42:45.240 & $-$07:01:10.20 & IRAS 18400$-$0704 & Feb 2016 & no detection  & C & - \\
ad3a-11836 & 18:49:10.392 & $-$06:53:03.48 & IRAS 18464$-$0656 & April 2016 & no detection & U & - \\
ad3a-12286 & 18:33:57.288 & $-$05:38:24.00 & no match        & March 2017 & no detection & - & - \\
ad3a-12346 & 18:41:13.464 & $-$05:26:30.84 & IRAS 18385$-$0529 & March 2017 & no detection & I & - \\
ad3a-12549 & 18:45:20.568 & $-$04:35:31.92 & IRAS 18427$-$0438 & March 2017 & no detection & - & -\\
ad3a-12574 & 18:40:51.912 & $-$04:31:38.28 & IRAS 18382$-$0434 & March 2017 & no detection & - & - \\
ad3a-12671 & 18:40:33.960 & $-$04:14:49.56 & IRAS 18379$-$0417 & March 2017 & no detection & C & - \\
ad3a-12827 & 18:34:40.272 & $-$03:50:13.92 & IRAS 18320$-$0352 & March 2017 & no detection & F & - \\
ad3a-12961 & 18:42:00.648 & $-$03:34:36.84 & no match        & March 2017 & no detection & - & - \\
ad3a-12989 & 18:44:01.392 & $-$03:31:27.48 & IRAS 18414$-$0334 & March 2017 & no detection & - & - \\
ad3a-13224 & 18:45:31.512 & $-$02:55:05.88 & IRAS 18428$-$0258 & March 2017 & no detection & - & - \\
ad3a-13287 & 18:47:50.952 & $-$02:38:12.12 & no match        & Dec 2015 & no detection & - & - \\
ad3a-13342 & 18:58:41.904 & $-$02:20:11.04 & IRAS 18560$-$0224 & March 2017 & no detection & - & - \\
ad3a-13348 & 18:44:20.688 & $-$02:18:47.52 & IRAS 18417$-$0221 & March 2017 & no detection & - & - \\
ad3a-13506 & 18:47:52.752 & $-$01:36:16.56 & IRAS 18452$-$0139 & March 2017 & no detection & I & - \\
ad3a-13722 & 18:52:16.536 & $-$00:31:32.52 & no match        & March 2017 & no detection & - & - \\
ad3a-13832 & 18:50:41.088 & $-$00:11:08.16 & IRAS 18481$-$0014 & March 2017 & no detection & F & - \\
ad3a-13872 & 18:51:44.208 & $-$00:05:07.44 & no match        & March 2017 & no detection & - & - \\
ad3a-13947 & 18:52:08.736 & +00:04:57.72 & IRAS 18495+0001 & March 2017 & no detection & - & - \\
ad3a-14048 & 18:50:12.168 & +00:20:44.52 & IRAS 18476+0017 & March 2017 & no detection & I & - \\
ad3a-14277 & 18:52:21.264 & +01:20:42.36 & IRAS 18498+0116 & March 2017 & no detection & - & - \\
ad3a-14335 & 18:49:19.056 & +01:35:46.32 & IRAS 18467+0132 & March 2017 & no detection & U & - \\
ad3a-14342 & 18:51:15.384 & +01:39:10.08 & IRAS 18487+0135 & March 2017 & no detection & C & - \\
ad3a-14511 & 18:51:24.912 & +02:45:22.32 & IRAS 18489+0241 & March 2016 & no detection & - & - \\
ae3a-00276 & 17:40:08.400 & $-$31:16:09.12 & no match        & March 2017 & no detection & - & - \\
ae3a-00349 & 18:47:32.592 & +00:46:21.00 & IRAS 18449+0042 & March 2016 & no detection & U & - \\
ce3a-00009 & 17:43:35.208 & $-$32:39:46.08 & IRAS 17403$-$3238 & Feb 2016 & no detection  & S & - \\
ce3a-00179 & 18:52:23.856 & $-$02:17:43.80 & IRAS 18498$-$0221 & March 2017 & no detection & U & - \\
ce3a-00185 & 18:49:25.536 & $-$00:39:25.20 & no match        & March 2017 & no detection & - & - \\
\enddata
\tablenotetext{c}{LRS class-A implies an O-rich AGB source which does not match the \cite{lewis20} IR-color classification. The source is very near the O/C boundary and uncertainties in the IR photometry may explain the disagreement (see Sect.\,\ref{con:redvar}). Additionally, thick-shelled AGB sources that would show sillicate features in absorption as opposed to emission may not fit the IR-color cut.}
\end{deluxetable*}


\par We confirm that C-bearing molecules can be detected in C-rich AGB sources within our covered frequencies and sensitivity, at least for this nearby source, but that a much improved sensitivity would be needed compared to the BAaDE survey or even this work in order to detect such lines at large.  
Additionally, no SiO masers were detected in any of the color-selected sources even though many maser lines were within our spectral and sensitivity range, as demonstrated by the detections in the O-rich objects.

\section{Discussion}\label{discussion}
Sources with IR colors indicative of both O-rich AGB stars and C-rich AGB stars were observed. Three out of five sources with IR-colors indicating O-rich stars show SiO maser lines, a forth source shows a tentative thermal SiO detection, and even the final source with no detection in this work can be classified as O-rich by its IRAS LRS. None of the 45 sources with IR-colors associated with C-rich stars (44 chosen by this criteria and one single-line source, \textit{ce3a-00142}, whose detection is still ambiguous) show definitive SiO maser lines. One source, \textit{ad3a-11235}, that was a C-candidate based on its IR colors and was known to be a carbon star from \cite{kvb97}, shows thermal HC$_3$N and SiO lines. As thermal SiO lines have been detected in several C-rich stars (\citealt{bujarrabal94, chau12}) and HC$_3$N is not expected in O-rich stars \citep{bujarrabal94}, these detections independently confirm this source as a C-rich AGB star. In general the candidate C-rich sources remain more ambiguous than the O-rich sources but the lack of SiO maser detections over multiple epochs along with a detection of HC$_3$N strongly suggests that the color-selection laid out in \citet{lewis20} is valid. 
Many arguments and data support the cut (Sect.\,\ref{pro:shortIR}$-$\ref{pro:nondetection}). We also discuss the uncertainty and limitations of the IR and radio data (Sect.\,\ref{con:redvar}$-$\ref{con:radio}). Finally sources with anomalously bright isotopologue lines and the serendipitous observation of their variability are discussed (Sect\,\ref{isovar}). 

\subsection{IR wavelength ranges}\label{pro:shortIR}
The success of the division relies on the addition of the 2MASS [K$_s$] information to the MSX data. 
It has been shown in several studies using multiple IR bands that mid-IR colors (eg. IRAS [12]$-$[25] or MSX [A]$-$[D]) are not effective in separating C- and O-rich AGB sources (\citealt{vdvH88}; \citealt{scc09}; \citealt{lewis20}).
\cite{vdvH88} show that the longer wavelength IR color [25]$-$[60] can divide O- and C-rich AGBs because C-rich AGBs are redder in these colors.
Here we have shown that C-rich AGBs are also redder in the shorter wavelength color [K$_s$]$-$[A].
This is in agreement with the general understanding of the SEDs of C- and O-rich AGB stars (see for example \citealt{ishihara11}), where the SEDs of these types of sources differ the most below 8 $\mu$m and above 20 $\mu$m. 
Cross-matching our sample of 51 sources to the IRAS database leads to 41 IRAS counter-parts. 
Thirty-one of these lack 60 $\mu$m information, making comparison between our color division and IRAS color-region classifications from \cite{vdvH88} uninformative. 
The division presented here is therefore particularly useful for sources that lack IRAS information (including but not limited to sources in crowded fields where 60 $\mu$m data is often poor quality) as they would otherwise be difficult to classify using IR colors.   

\subsection{Agreement with IRAS LRS}\label{pro:lrs}
\begin{figure}[htb!]
    \centering
    \includegraphics[trim=0cm 0cm 0cm .1cm, clip=true, width=.45\textwidth]{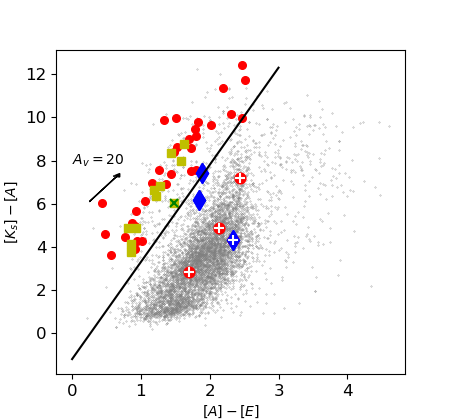}
    \caption{Color-color diagram. Colored symbols represent the follow-up sample and are categorized based on the \cite{kvb97} analysis of IRAS LRS, while grey dots represent BAaDE sources not observed in the follow-up program. Yellow squares are identified as C-type meaning they show the 11.2 $\mu$m SiC feature or F-type meaning their spectra are featureless (i.e., they are likely C-rich AGB sources), blue diamonds are identified as A- or E-type meaning they show the 9.7 $\mu$m silicate feature in either absorption or emission (i.e., they are likely O-rich AGB sources), and red dots are sources without a LRS identification or designations that do not aid in O-/C-rich classification. Sources with SiO-only detections in the follow-up spectra are marked with a white plus and correspond to either unidentified or O-rich LRS sources. \textit{ad3a-11235} which shows thermal HC$_3$N and SiO in its spectra and is identified as C-rich by its LRS is marked as a green cross. The reddening vector corresponding to 20 magnitudes of visual extinction \citep{lumsden02}, is shown as the black arrow.}
    \label{fig:color-color}
\end{figure}

Because of the lack of definitive C-bearing molecular detections in many of the potential C-rich AGB stars, we also match our sample to IRAS LRS classifications where applicable. 
We use the categorizations by \cite{kvb97}, and consider LRS types A and E (9.7 $\mu$m feature in absorption and emission respectively) to be O-rich AGBs and LRS types C and F (11.2 $\mu$m feature and featureless spectra respectively) to be C-rich \citep{volk92}.
Figure \ref{fig:color-color} shows LRS identified C-rich and O-rich AGB sources from our sample on a MSX 2MASS color-color diagram similar to Fig.\,\ref{fig:selection}. 
Also marked are the sources with spectral detections from this work and the O/C-dividing line described by Equation 1. The distribution of C- and O-rich objects follows the color cut very well; therefore, both IRAS LRS and our own independent results support the color division. 

\subsection{Lack of detections over multiple epochs}\label{pro:nondetection}
Neither a nondetection in the BAaDE observations nor in the new observations necessarily implies a C-rich AGB, as some O-rich AGBs do not host maser emission, SiO maser emission is variable, and other types of IR sources could contaminate our sample. 
However, given that all but two (\textit{ad3a-11235} and \textit{ce3a-00142}) of the sources with colors indicative of C-rich AGB sources have featureless spectra at 43 GHz over at least two separate epochs (and none show definitive SiO masers), we can effectively rule out that this is a sample of SiO-maser sources. 
Based on redetection rates of known BAaDE masers (\citealt{ylva}; \citealt{michael18}), SiO masers are detectable within the BAaDE sensitivity during about 80\% of the stellar cycle. The odds of randomly observing an SiO maser source twice while it is off (with the observations separated significantly in time) are therefore quite low at 4\% (20\% of 20\%), and it is nearly impossible ($\sim$10$^{-60}$\%) to observe 44 maser-bearing sources coincidentally in this phase twice. If these sources are a population with similar characteristics, then we can be certain that they are not a population of O-rich AGB sources with circumstellar conditions conducive to maser emission. 
More generally, the nondetections reported here, the detection of HC$_3$N in \textit{ad3a-11235}, and distribution of LRS sources reported in this work add to the evidence given in \cite{lewis20} and continue to support the conclusion that thin-shelled C-rich AGB sources lie above the line in Fig.\,\ref{fig:color-color}. 

\subsection{Potential effects of reddening and variability}\label{con:redvar}
Two primary effects could change the apparent color of a source to position it on the wrong side of the division between O-rich and C-rich AGB sources: reddening and variability. Although the sources observed here lie close to the Galactic plane and no reddening corrections have been applied (for a number of reasons including, but not limited to, a lack of distance estimates), reddening is probably not the main source of error in identifying an individual source as C- or O-rich. This is because the reddening vector and the O/C division are roughly parallel (see Fig.\,\ref{fig:color-color}). Sources whose apparent colors are strongly affected by reddening through interstellar dust will have intrinsic colors that slide towards the bluer part of the diagram as shown by the slope of the reddening vector, and therefore are likely to stay on the same side of the O/C division even if large corrections are applied. As seen in the figure, even if sources were corrected for 20 magnitudes of visual extinction, consistent for sources near the Galactic center, they would still almost all stay on their respective sides of the division. 
\par A greater source of error is the non-simultaneity of the the 2MASS [K$_s$] and the MSX photometry. Our sources are variable AGB stars whose IR magnitudes can change by more than a magnitude from maximum to minimum, and the sources could have been in different parts of their stellar phases during the 2MASS and MSX measurements. This mostly affects the [K$_s$]$-$[A] color because it includes measurements from both surveys. At worst, if these measurements were made completely out of phase, this could misrepresent the [K$_s$]$-$[A] color by about two magnitudes, which can indeed cause sources to appear to be on the other side of the division. This source of error can move sources in both directions across the division and should in general not have a large effect on a statistical sample, although individual sources, like \textit{ad3a-09711} whose LRS shows a silicate absorption feature but is on the C-side of the division, can be affected. 
\subsection{Sensitivity limit of radio data}\label{con:radio}
We detect emission in only one source that was chosen based on color, \textit{ad3a-11235}. This source has been confirmed as a C-rich AGB source through several methods (\citealt{MSIII}, \citealt{kvb97}, \citealt{alksnis}, \citealt{chen}) and is likely less than a kpc away (see Sect.\,\ref{results:color}). It is encouraging that firstly, the color criteria is picking out known C-rich AGB sources and secondly, that a C-bearing molecular detection can be made within these sources within our frequency coverage. However, as there is only a single detection made in any source chosen by this criteria, this is not an efficient spectroscopic method for definitively identifying C-rich AGB sources for large samples. This is due to the relative weakness of the thermal C-bearing lines available in this frequency range. A much improved sensitivity would be required to positively confirm a large number of C-rich AGB sources in this frequency range. \par Furthermore, the selection criteria for this sample included a magnitude cut intended to limit the distances of the sources and increase the chances of molecular detections. This criteria, which was imposed because of the sensitivity limit of the radio data, implies that our sample is limited to stars in the Galactic disk. These data therefore do not exclude the possibility that the O/C division could have a metallicity dependency that would alter the division in other environments, such as the Galactic bulge. As the division was initially proposed based on SiO detection rates from across both the disk and the bulge from the BAaDE survey, we find it unlikely that the effect of metallicity on the division is significant. 
\subsection{Variability of isotopologue lines}\label{isovar}
\begin{figure}[htb]
    \centering
    \includegraphics[width=.45\textwidth]{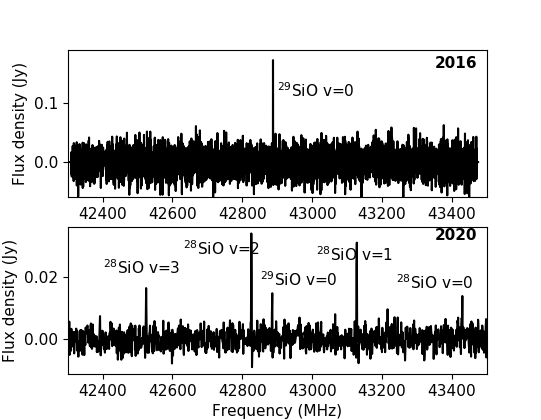}
    \caption{Spectra for the source \textit{ad3a-01808} showing that the line ratios between $^{28}$SiO and $^{29}$SiO transitions can completely reverse between observations. Top: data from the 2016 portion of the BAaDE survey. Bottom: data from the 2020 follow-up observations limited in frequency range to match the earlier observations. Note that the flux density scale is different in the top and bottom panels due to the improved sensitivity in the new observations. None of the follow-up detections would be significant at the sensitivity of the BAaDE data. }
    \label{fig:ad3a-01808}
\end{figure}
The ambiguity of the initial BAaDE single-line detection in the single-line sources described in Sect.\,\ref{sourcesline} was resolved by reobserving to a higher sensitivity and detecting additional lines in the same spectra simultaneously. These sources show SiO maser emission and are O-rich AGB stars as their 2MASS and MSX colors suggest. This means that during the BAaDE survey observations, the brightest line in the 43 GHz spectra of these sources was the $^{29}$SiO v=0 line. This was unexpected as the $^{28}$SiO v=1 and/or v=2 lines are much more common and usually dominate over the $^{29}$SiO line by several factors. Observations of these sources over two epochs show that, not only can $^{29}$SiO v=0 be the brightest, and indeed only, line detected in a BAaDE source, but that the ratios between the $^{29}$SiO and $^{28}$SiO lines are highly time-dependent. The line ratios in a source where the $^{29}$SiO v=0 line is the brightest can completely reverse within the span of a few years, so that the source displays typical line ratios at a later epoch. Source \textit{ad3a-01808} shows the variability of the $^{29}$SiO and $^{28}$SiO line ratios (see Fig.\,\ref{fig:ad3a-01808}). In this source, the only detectable line in the BAaDE data was a transition near 42.9 GHz, while the follow-up data revealed five SiO lines three of which were brighter than the initially detected 42.9 GHz line---which is now identified to be the $^{29}$SiO v=0 line. 
\par Both the fact that the $^{29}$SiO v=0 can be brighter than the $^{28}$SiO v=1 and v=2 lines and the fact that these atypical line ratios can revert are newly reported observations that have yet to be incorporated into, or explained by, maser pumping models. It seems unlikely that anomalously bright $^{29}$SiO lines can be caused simply by anomalously high abundances of $^{29}$SiO as these bright $^{29}$SiO sources can switch to typical line ratios in a matter of years. This precludes using maser detections like this as conclusive identifiers of Thorne-\.Zytkov objects \citep{para95}, at least until the anomalous brightness is better understood.

\section{Conclusions}\label{conclusion}
Two epochs of 43 GHz observations of thin-shelled AGB sources confirm that the O-/C-rich AGB division from \cite{lewis20} is very effective. Following up on single-line detections from the BAaDE survey leads mainly to more positive confirmations of O-rich objects through the detection of SiO emission, while following up on sources whose IR colors suggest they are C-rich leads to mostly nondetections and one instance of a detection of a C-bearing molecule. The large number of nondetections among the sources chosen by color supports the color cut given the typical strength and prevalence of SiO masers in the frequencies covered. Deep integrations at mm or IR wavelengths are likely necessary to directly spectroscopically confirm C-rich AGB sources from the BAaDE survey as the limited sensitivity of the observations presented here only show detectable emission from a single, relatively nearby source. The carbon-bearing molecular lines targeted by these observations are too weak to be commonly detected in a reasonable amount of time. Identifications from IRAS LRS observations made by \cite{kvb97} also fit the color division well. Our photometric classification is more widely applicable than IRAS LRS in the Galactic plane because many Galactic IR sources lack LRS or even reliable IRAS photometry, especially at longer wavelengths, due to confusion.  
\par Finally, the $^{29}$SiO v=0 line can be brighter than the traditionally dominant $^{28}$SiO v=1 and v=2 lines. This reversal of line ratios changes over time and can completely switch in the span of $\sim$3 years suggesting that the abnormal ratios are not caused by an abundance effect. 

\section*{Acknowledgments}
The National Radio Astronomy Observatory is a facility of the National Science Foundation operated under cooperative agreement by Associated Universities, Inc. 
Support for this work was provided by the NSF through the Grote Reber Fellowship Program administered by Associated Universities, Inc./National Radio Astronomy Observatory.
This paper makes use of the VLA dataset 19B-219.\\
This research has made use of the SIMBAD database, operated at CDS, Strasbourg, France.\\
This research has made use of the NASA/IPAC Infrared Science Archive, which is funded by the National Aeronautics and Space Administration and operated by the California Institute of Technology.
This research made use of data products from the Midcourse Space Experiment. Processing of the data was funded by the Ballistic Missile Defense Organization with additional support from NASA Office of Space Science.
This publication makes use of data products from the Two Micron All Sky Survey, which is a joint project of the University of Massachusetts and the Infrared Processing and Analysis Center/California Institute of Technology, funded by the National Aeronautics and Space Administration and the National Science Foundation.

%

\facilities{VLA, MSX, 2MASS, IRAS, IRSA}





\appendix
\vspace{-0.7cm}
\section{Spectra}
\begin{figure*}[htb!]
    \centering
    \includegraphics[width=.48\textwidth]{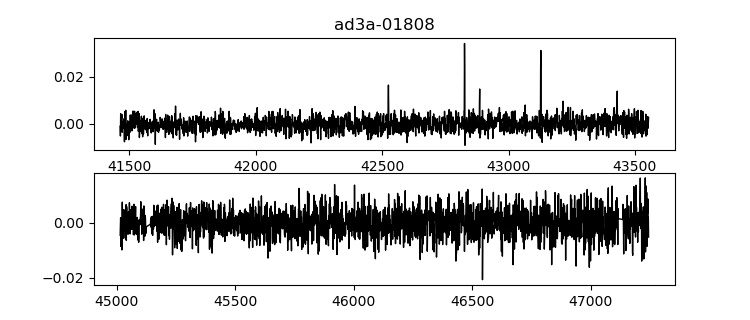}
    \includegraphics[width=.48\textwidth]{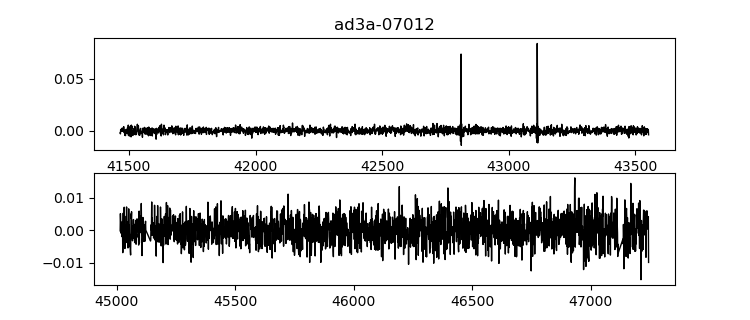}
    \includegraphics[width=.48\textwidth]{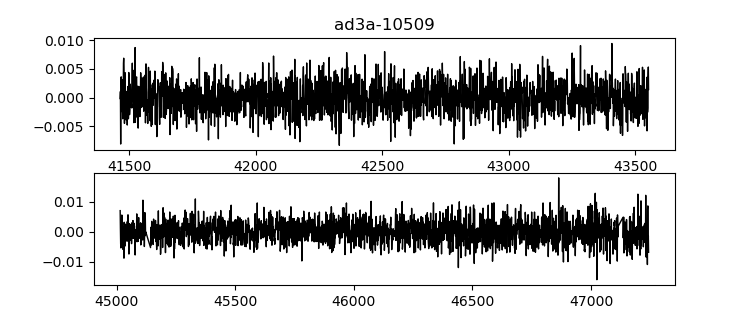}
    \includegraphics[width=.48\textwidth]{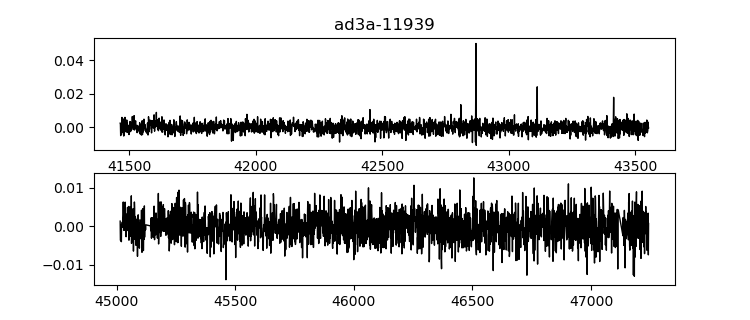}
    \includegraphics[width=.48\textwidth]{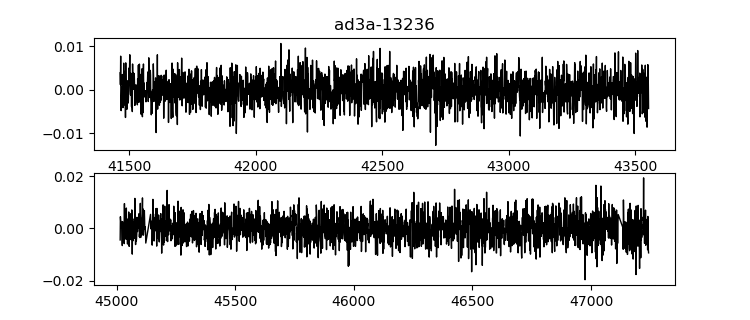}
    \includegraphics[width=.48\textwidth]{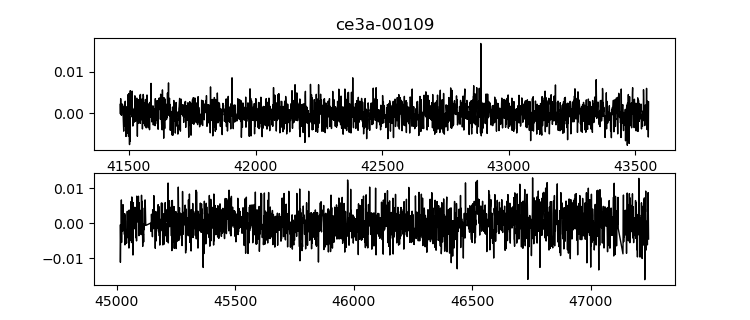}
    \includegraphics[width=.48\textwidth]{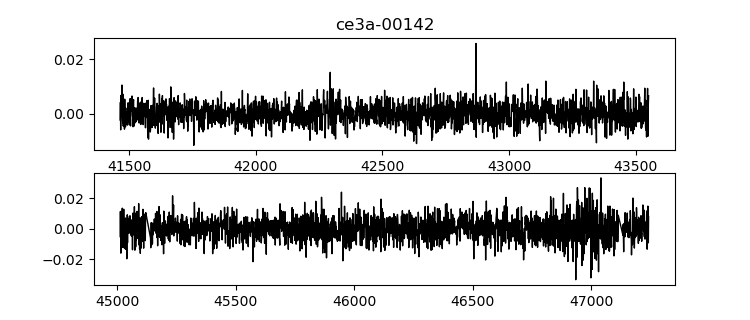}
    \caption{All single-line-selected follow-up spectra (Table \ref{tab:singlines}). Frequencies (horizontal-axes) are in MHz and flux densities (vertical-axes) are in Jy.}
    \label{fig:singlelines}
\end{figure*}

This appendix contains all 51 of the 2020 VLA spectra labeled by BAaDE source name. The seven sources chosen by their BAaDE spectra (Table \ref{tab:singlines}) are displayed in Fig.\,\ref{fig:singlelines} and the 44 sources chosen by their color (Table \ref{tab:ccolor}) are in Fig.\,\ref{fig:00345}. Each spectra consists of two panels, one centered near 42.5 GHz (top panel) and the other centered near 46 GHz (bottom panel). No channel averaging is applied.

\newpage
\begin{figure*}[p!]
    \centering
    \includegraphics[width=.48\textwidth]{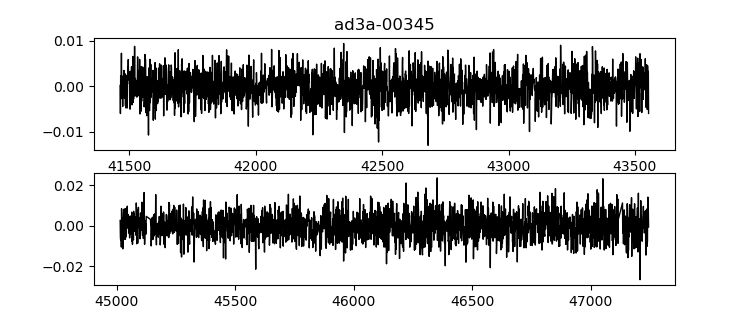}
    \includegraphics[width=.48\textwidth]{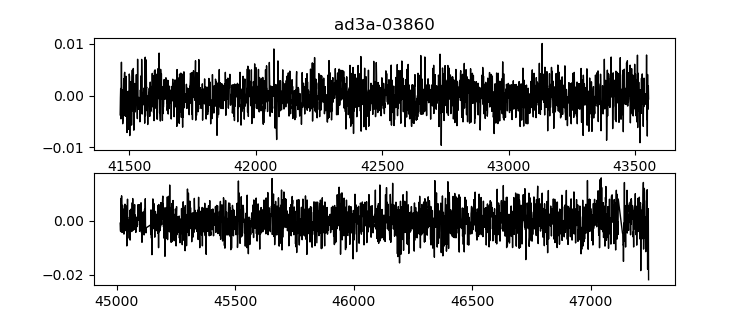}
    \includegraphics[width=.48\textwidth]{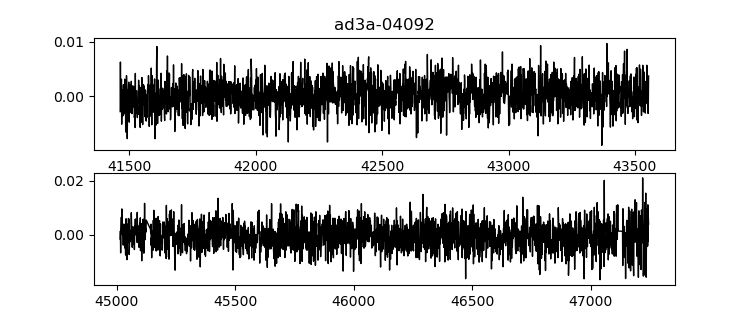}
    \includegraphics[width=.48\textwidth]{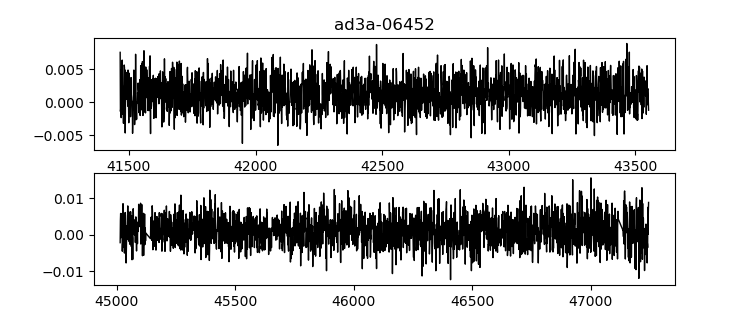}
    \includegraphics[width=.48\textwidth]{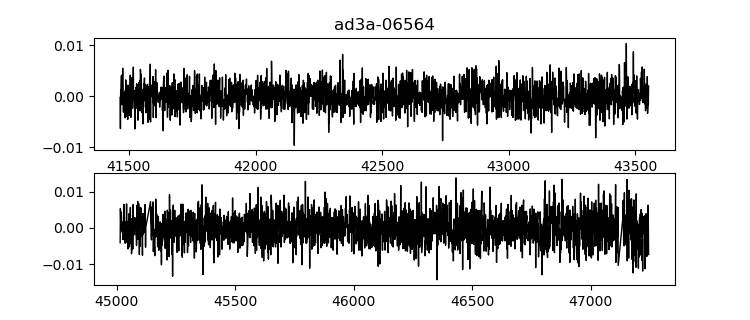}
    \includegraphics[width=.48\textwidth]{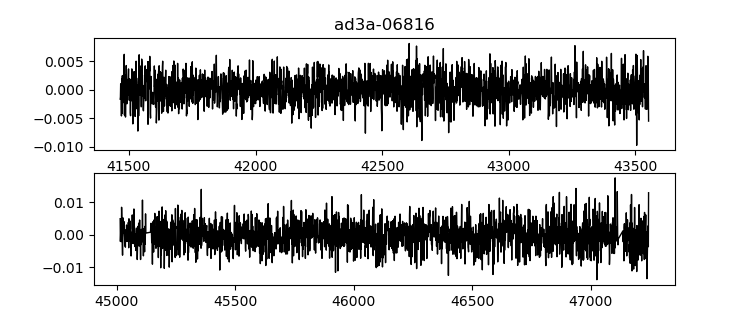}
    \includegraphics[width=.48\textwidth]{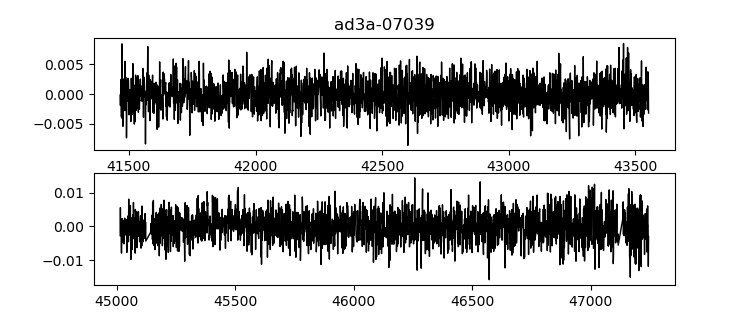}
    \includegraphics[width=.48\textwidth]{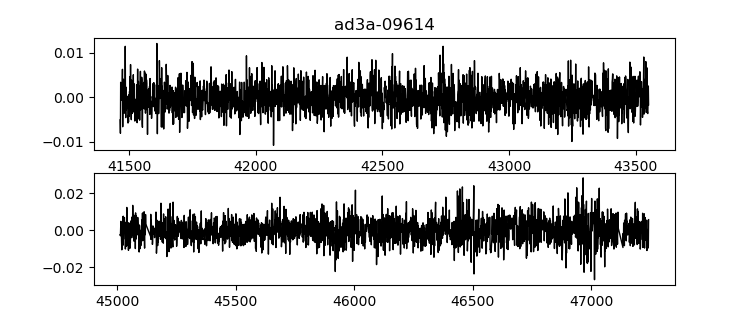}
    \includegraphics[width=.48\textwidth]{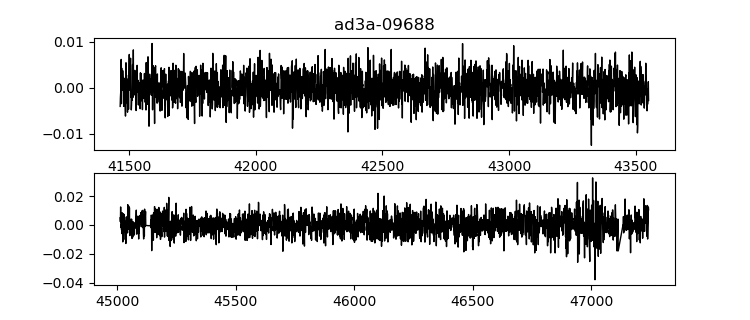}
    \includegraphics[width=.48\textwidth]{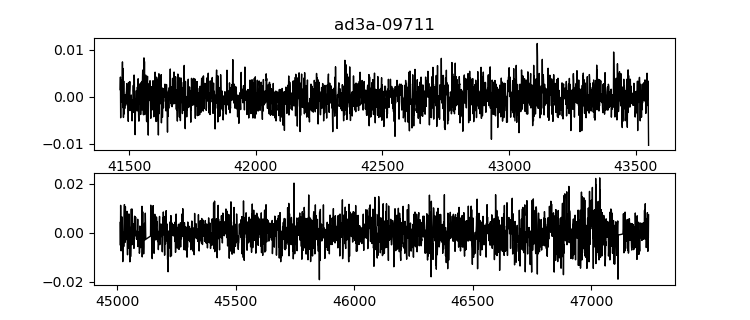}
    \caption{All color-selected follow-up spectra (Table \ref{tab:ccolor}). Frequencies (horizontal-axes) are in MHz and flux densities (vertical-axes) are in Jy.}
    \label{fig:00345}
\end{figure*}\clearpage

\newpage
\begin{figure*}[p!]
    \addtocounter{figure}{-1}
    \centering
    \includegraphics[width=.48\textwidth]{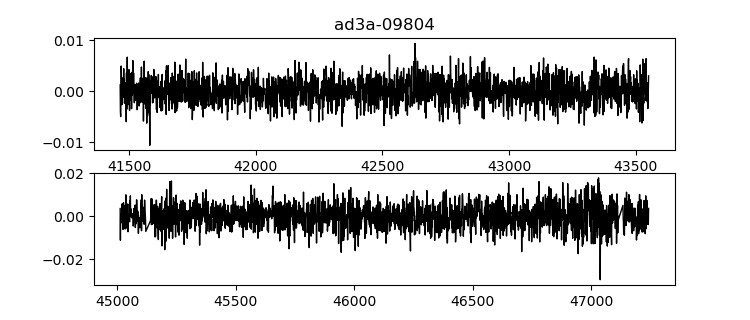}
    \includegraphics[width=.48\textwidth]{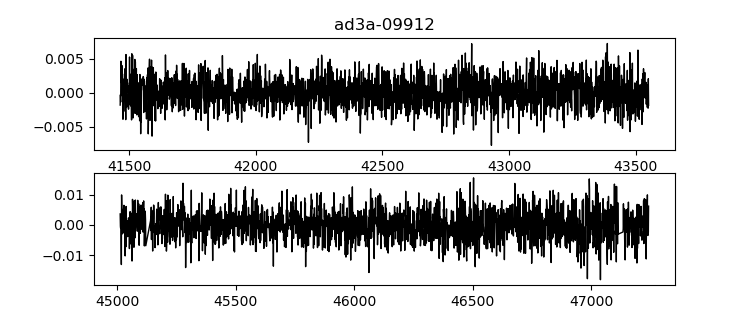}
    \includegraphics[width=.48\textwidth]{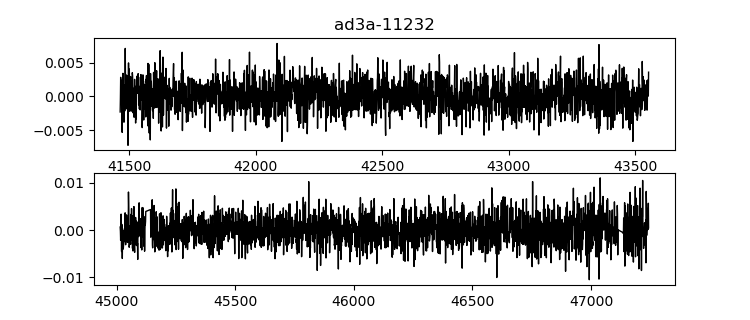}
    \includegraphics[width=.48\textwidth]{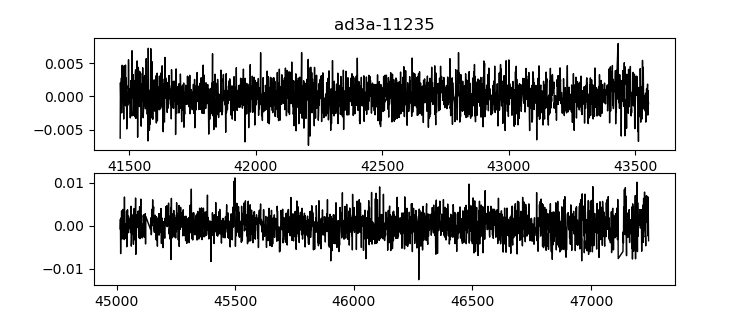}
    \includegraphics[width=.48\textwidth]{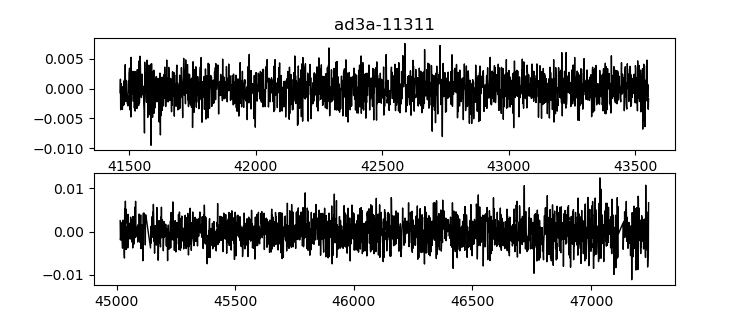}
    \includegraphics[width=.48\textwidth]{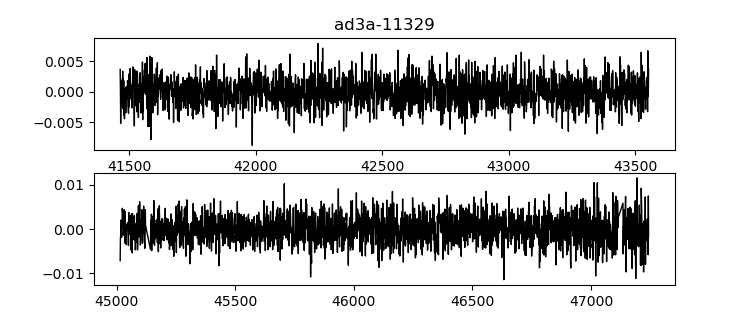}
    \includegraphics[width=.48\textwidth]{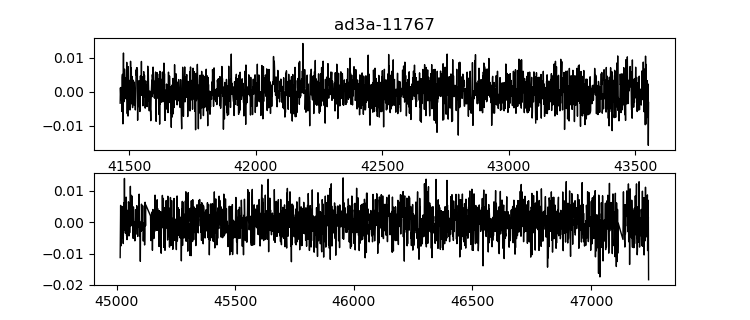}
    \includegraphics[width=.48\textwidth]{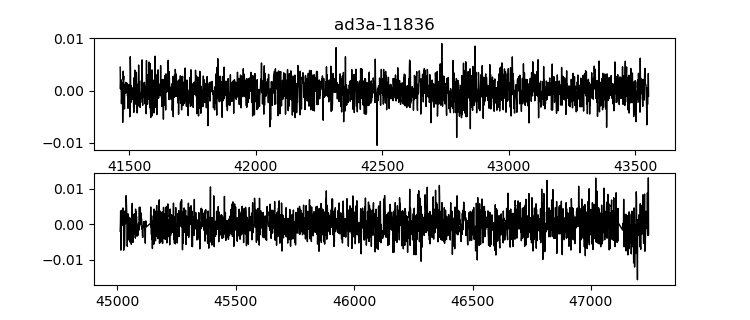}
    \includegraphics[width=.48\textwidth]{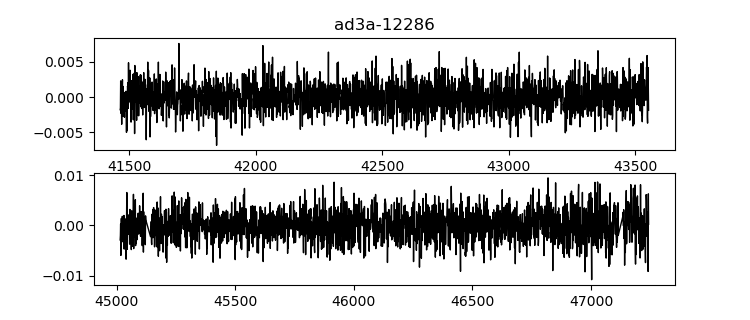}
    \includegraphics[width=.48\textwidth]{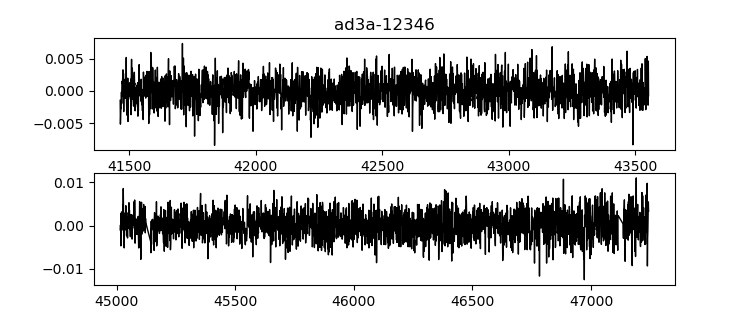}
    \caption{cont.}
    \label{fig:11232}
\end{figure*}\clearpage

\newpage
\begin{figure*}[p!]
    \addtocounter{figure}{-1}
    \centering
    \includegraphics[width=.48\textwidth]{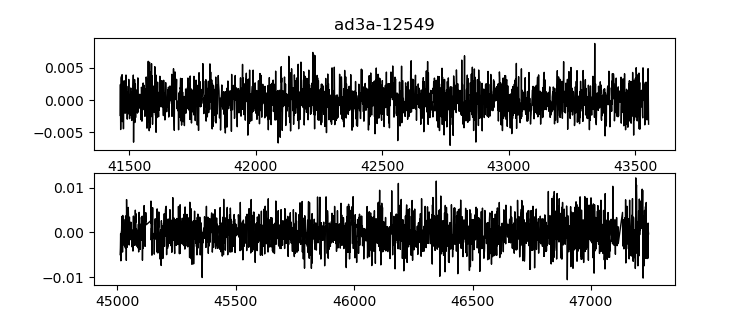}
    \includegraphics[width=.48\textwidth]{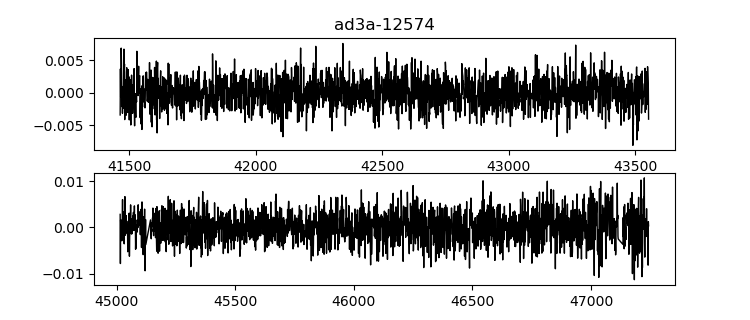}
    \includegraphics[width=.48\textwidth]{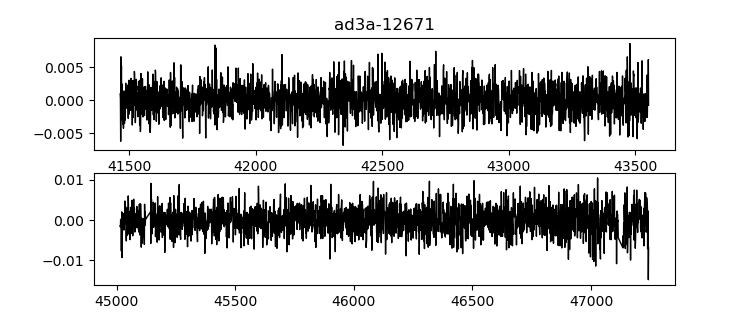}
    \includegraphics[width=.48\textwidth]{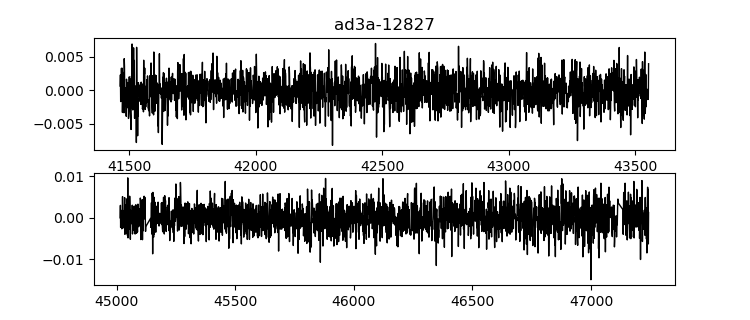}
    \includegraphics[width=.48\textwidth]{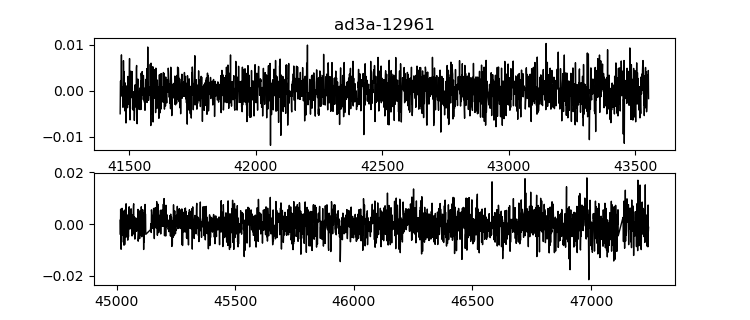}
    \includegraphics[width=.48\textwidth]{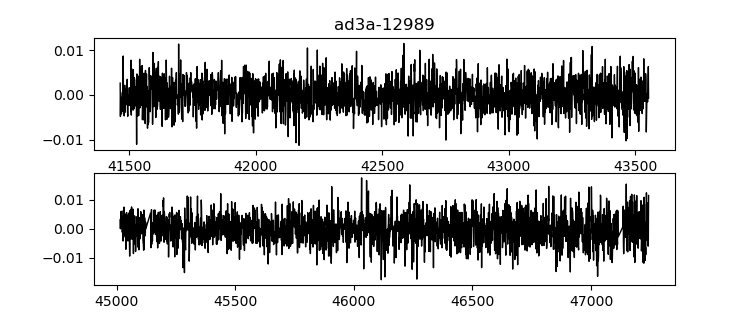}
    \includegraphics[width=.48\textwidth]{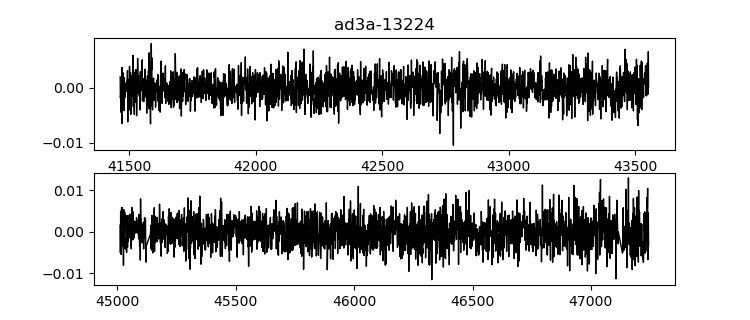}
    \includegraphics[width=.48\textwidth]{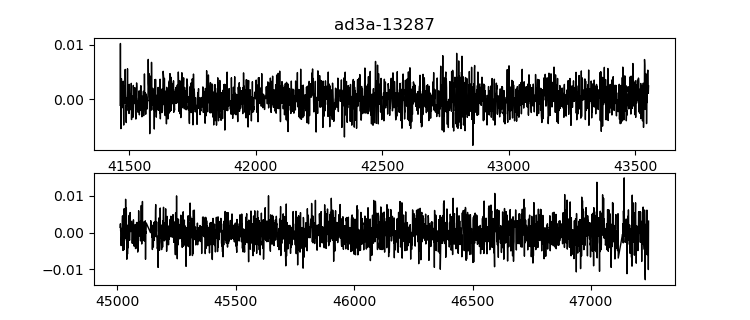}  
    \includegraphics[width=.48\textwidth]{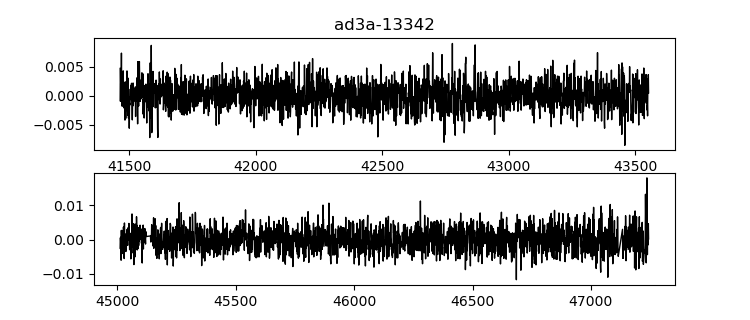}
    \includegraphics[width=.48\textwidth]{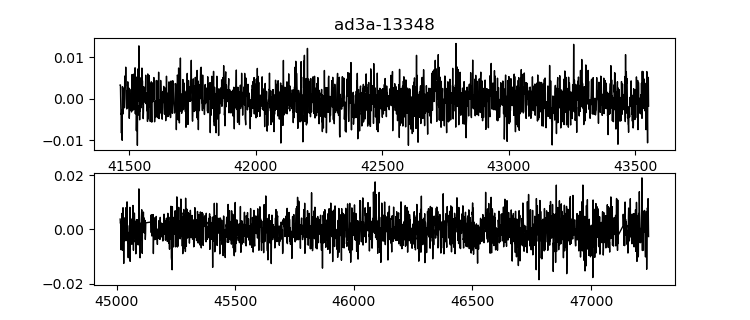}
    \caption{cont.}
    \label{fig:12549}
\end{figure*}\clearpage

\newpage
\begin{figure*}[p!]
    \addtocounter{figure}{-1}
    \centering
    \includegraphics[width=.48\textwidth]{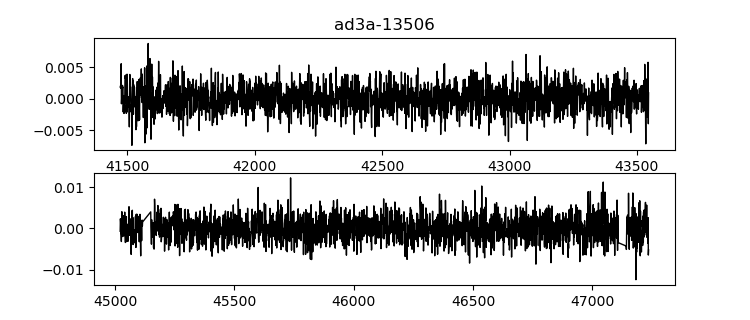}
    \includegraphics[width=.48\textwidth]{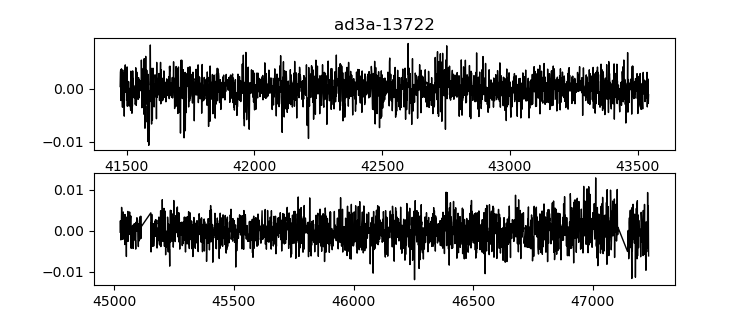}
    \includegraphics[width=.48\textwidth]{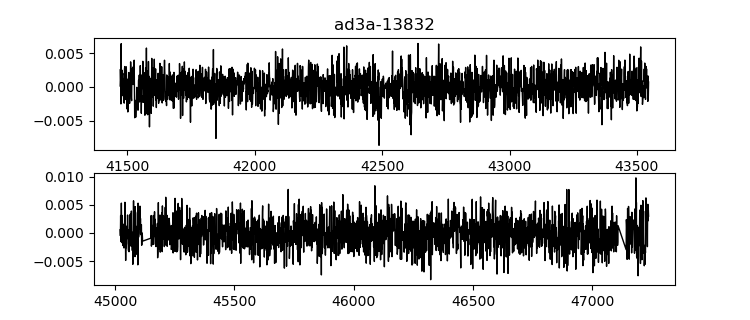}
    \includegraphics[width=.48\textwidth]{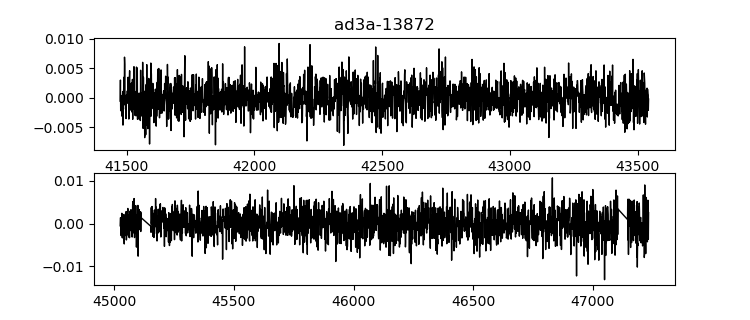}
    \includegraphics[width=.48\textwidth]{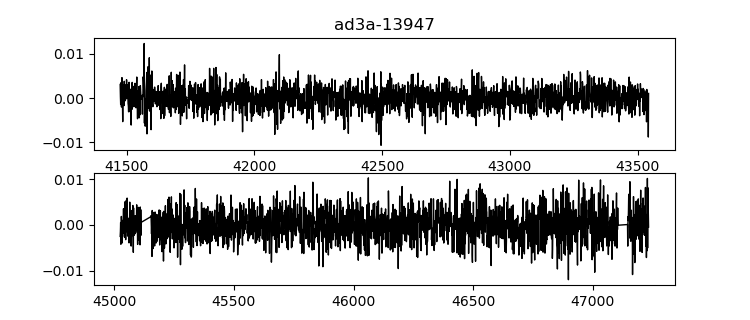}
    \includegraphics[width=.48\textwidth]{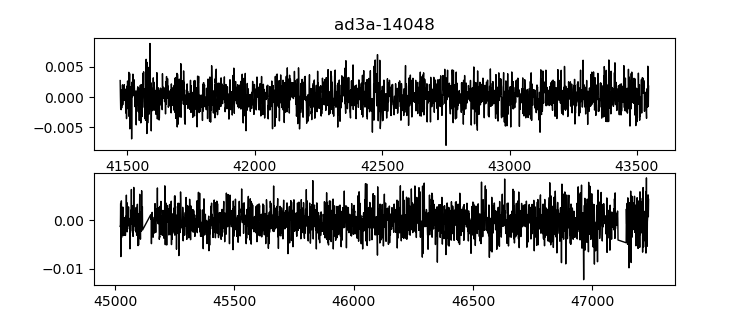}
    \includegraphics[width=.48\textwidth]{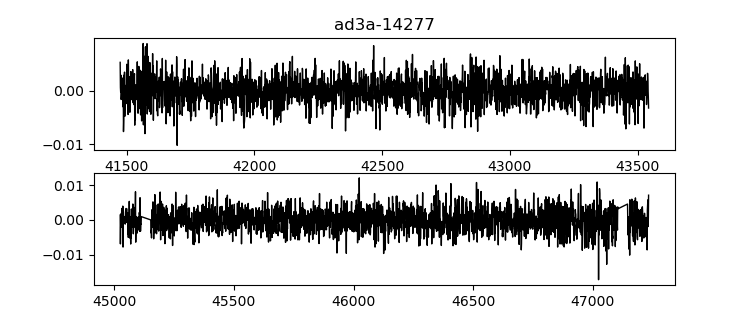}
    \includegraphics[width=.48\textwidth]{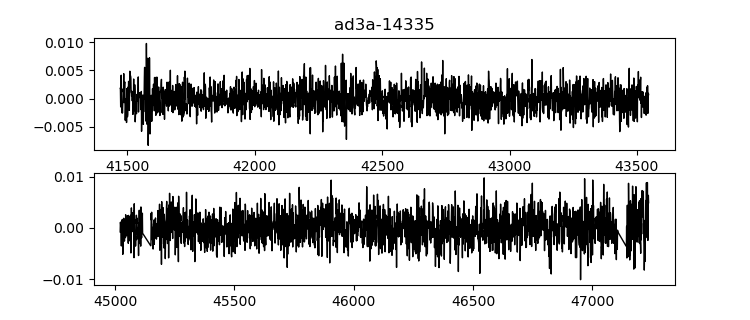}
    \includegraphics[width=.48\textwidth]{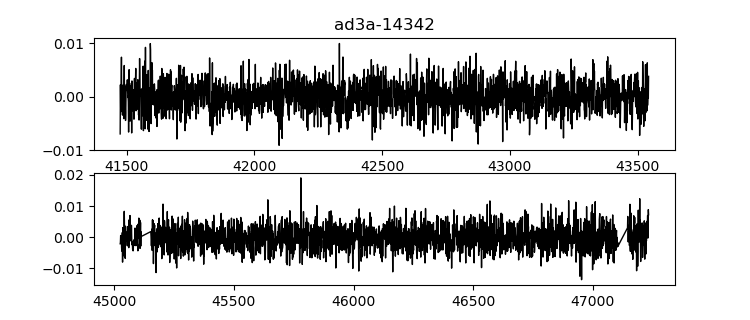}
    \includegraphics[width=.48\textwidth]{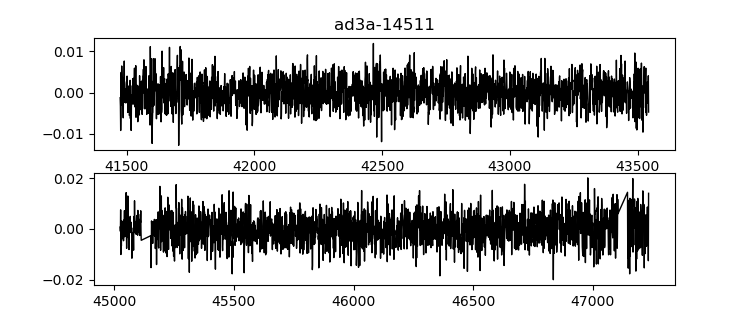}
    \caption{cont.}
    \label{fig:13722}
\end{figure*}\clearpage

\newpage
\begin{figure*}[ht!]
    \addtocounter{figure}{-1}
    \centering
    \includegraphics[width=.48\textwidth]{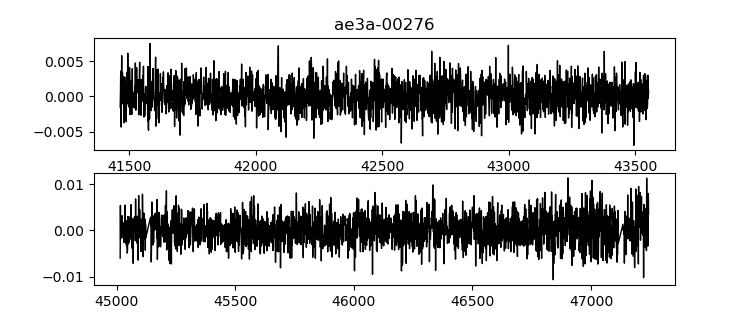}
    \includegraphics[width=.48\textwidth]{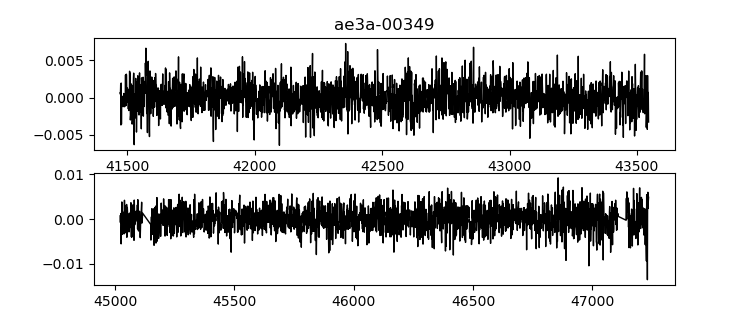}
    \includegraphics[width=.48\textwidth]{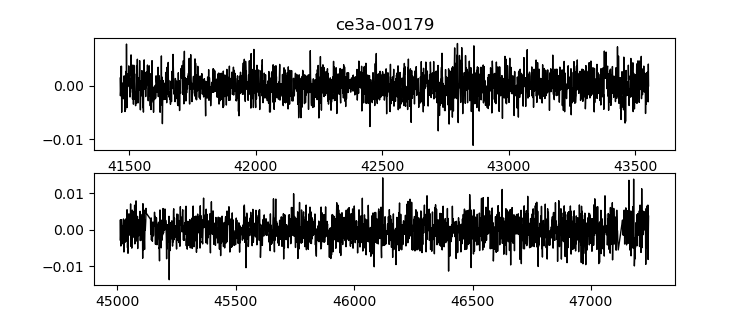}
    \includegraphics[width=.48\textwidth]{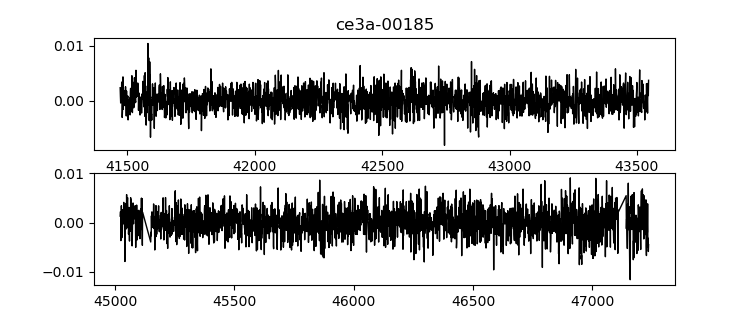}
    \caption{cont.}
    \label{fig:ce}
\end{figure*}

\bibliography{sample63}{}
\bibliographystyle{aasjournal}



\end{document}